\documentclass[aps,prb,reprint,superscriptaddress,showpacs,floatfix,10pt]{revtex4-1}
\usepackage{graphicx}
\usepackage{hyperref}

\begin{document}

\title{Number of observable features in the acoustic-Raman spectra of nanocrystals}
\author{Lucien Saviot}
\email{lucien.saviot@u-bourgogne.fr}
\affiliation{Laboratoire Interdisciplinaire Carnot de Bourgogne,
UMR 6303 CNRS-Universit\'e de Bourgogne,
9 Av. A. Savary, BP 47 870, F-21078 Dijon Cedex, France}

\author{Nicolas Combe}
\author{Adnen Mlayah}
\affiliation{Centre d'Elaboration de Mat\'eriaux et d'Etudes
Structurales, CNRS UPR 8011, 29 rue J. Marvig, BP 94347, 31055 Toulouse cedex
4, France}
\affiliation{Universit\'e de Toulouse ; UPS ; F-31055 Toulouse, France}

\begin{abstract}
Low-frequency Raman scattering spectra are presented for gold nanocrystals with diameters 3.5 and 13~nm.
The frequencies of the Raman peaks but also their number are shown to vary with the nanocrystal size.
These results are analyzed using both the continuous elastic medium approximation and an atomistic approach.
We show that the number of atoms in the nanocrystal
determines an upper limit of the number of observable Raman features.
The frequency range in which the continuous elastic medium approximation is valid is defined by comparison with the calculations based on the atomistic approach. 
\end{abstract}

\maketitle

\section{Introduction}

Many advances have been made in the last decades regarding the experimental investigations
and theoretical descriptions of confined acoustic vibrations in nanocrystals (NCs).
In particular, many works have been devoted to inelastic light scattering spectra (Raman or Brillouin) which are known to depend on the size, shape, composition, environment and  lattice structure of the NCs.\cite{MlayahLSS07,SaviotHNP10}
Recent papers have reported a variety of spectra for very small
(a few nanometers) to quite large (hundreds of nanometers) NCs, especially for metallic NCs.
For NCs sizes smaller than the optical wavelength, the Raman selection rules dramatically restrict the number of Raman active vibration modes.\cite{duval92}
For larger nanoparticles (hundreds of nanometers), several vibration modes can contribute to the Raman/Brillouin spectra.\cite{MontagnaPRB08}
However, while the assignment of the spectral features has received a lot of attention, the question of their number remains largely unaddressed.

Assuming a continuous and isotropic medium, the vibration eigenmodes of a spherical particle can be derived
using Lamb's model.\cite{lamb1882}
One obtains spheroidal and torsional modes here 
labeled $S_\ell^ n$ and $T_\ell^n$, respectively; $\ell$ and $n$ being the angular momentum and overtone index.
The number of vibration eigenmodes is infinite. 
This prediction is however bound to fail for clusters consisting of few atoms.

In this work, we present experimental results obtained for gold NCs which demonstrate that the number of acoustic vibration modes and associated Raman peaks strongly depend on the NC size, \textit{i.e.}, on the number of atoms in the NC.
Assuming a continuous elastic medium, we derive an upper limit for the number of observable Raman peaks.
In order to properly describe the high frequency vibration modes with wavelengths
comparable to the interatomic distance, we also use an atomistic description.
In this approach, the non-linearity of the acoustic phonon dispersions as well as the contribution of surface atoms to the vibrational dynamics are taken into account.

\section{Results and discussion}

\subsection{Raman setup, samples and spectra}

Low-frequency Raman scattering was measured on approximately spherical 13 and 3.5~nm gold
NCs, chemically synthesized in colloidal solutions.
In all Raman experiments the Argon/Krypton ion laser intensity was around
0.5~mW/$\mu$m$^2$ at the focal plane.
The scattered light was collected through a 50$\times$
objective of a confocal microscope with a 0.8 numerical aperture and
dispersed using a Jobin-Yvon T64000 spectrometer which allows for high rejection of the
Rayleigh scattering.
The laser spot size is about 10~$\mu$m$^2$ and the
incidence angle is 60$^\circ$ with respect to the normal of the sample surface.

\begin{figure}
\includegraphics[width=\columnwidth]{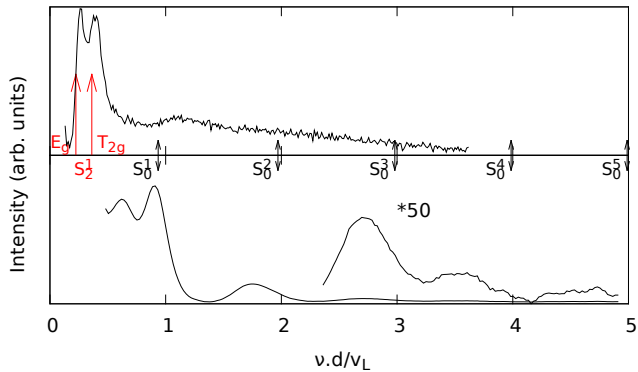}
\caption{\label{exp}(Color online)
Low-frequency Raman spectra of gold NCs with diameters of 3.5 (top) and 13~nm (bottom).
The Raman shift $\nu$ has been multiplied by the diameter $d$ and divided by the average longitudinal sound velocity $v_L$.
The arrows indicate the spectral features due to
$S_2^1$ (large arrows, red online) with $E_g$ and $T_2g$   and
$S_0^n$ vibrations (small arrows, black online) with $n=1, \ldots 5$ from left to right.}
\end{figure}

In order to enhance the Raman scattering from the 13~nm NCs, the latter were
coated on gold nanodisks and the solvent was evaporated.
The NCs were then excited at 647.1~nm in resonance with
the surface plasmons of the nanodisks as described in
Ref.~\onlinecite{TripathyNL11}.
The nanodisks provide a spectrally sharp surface plasmon resonance associated with spatially localized electric near-fields which allows for a strong enhancement of the Raman scattering.
As shown in Fig.~\ref{exp} several overtones of the NCs vibration modes are observed.\cite{TripathyNL11}
Three additional low-frequency peaks are present but not resolved in this spectrum due to the Rayleigh scattering.
They have been reported elsewhere and assigned to the $S_2^1$ (split into $E_g+T_{2g}$ by elastic anisotropy) and $S_2^2$ vibrations.\cite{AdichtchevPRB09}

The Raman spectrum of the 3.5 nm NCs is also shown in Fig.~\ref{exp}.
The NCs were coated on a 300~nm SiO$_2$/Si substrate and excited at 520~nm in
resonance with both the surface plasmons of the nanoparticles and one of the SiO$_2$/Si optical cavity modes in order to take
advantage of a double resonance effect: surface plasmons and interference enhanced Raman scattering.\cite{nemanich80}
This is particularly useful in the case of small particles with a very weak optical density.
Because of the smaller size, the lowest frequency peaks are now observed ($S_2^1$ split into $E_g+T_{2g}$ modes) together with the $S_0^1$ band. This modes assignement is based on the Raman selection rules (breathing modes generate only polarized Raman scattering whereas quadrupolar modes give rise to both polarized and depolarized scattering). 

In Fig.~\ref{exp}, the Raman shift has been multiplied by the diameter $d$ of the NCs in order
to compare the number of spectral features for different NC sizes.
In addition, it has been normalized to the longitudinal sound velocity of gold. In that way, the frequencies of the breathing vibration modes (calculated using the continuous elastic medium approximation)
are located approximately at integer abscissa (short arrows in Fig.~\ref{exp}).
This figure clearly shows that changing the size of the NCs significantly changes the number of Raman features: the Raman spectrum of the 13~nm NCs exhibits more spectral features than that of the 3.5~nm NCs.

\subsection{Lamb's model}

We first discuss the vibrational dynamics of spherical NCs in the frame of Lamb's model.\cite{lamb1882}
In this model, the vibrating medium is assumed to be continuous and isotropic.
The vibration eigenmodes are obtained from the Navier-Stokes hydrodynamic equations using $v_L=3330$ and $v_t=1250$~m/s for the longitudinal and transverse sound speeds of gold.\cite{SaviotPRB09}
According to the Raman selection rules,\cite{duval92} for NC sizes small compared to the optical wavelength, 
only spheroidal vibration modes with angular intergers $\ell=0$ and $\ell=2$ (i.e., $S_0^n$ and $S_2^n$ modes) are Raman active in agreement with experiments.\cite{AdichtchevPRB09}
In this work, we will focus on the existence of breathing vibration modes $S_0^n$ and related Raman features.

It is well-known\cite{TamuraJPC82} that the continuous elastic medium approximation becomes questionable when the NC contains only few atoms and/or the wavelength of the vibration mode is comparable to the interatomic distance. 
In order to take into account these limitations, 
a simple approach consists in considering only the $3N$ lowest-frequency vibration modes ($N$ being the number of atoms in the NC) with angular periodicity larger than the interatomic distance ($\ell_{max}=\pi\frac{d}{a \sqrt{2}}-\frac12$).\cite{TamuraJPC82}
This simple rule gives a rough estimation of the number $N_{breathing}$ of allowed vibration modes and their frequency range $\nu_{max}$.

\begin{figure}
\includegraphics[width=\columnwidth]{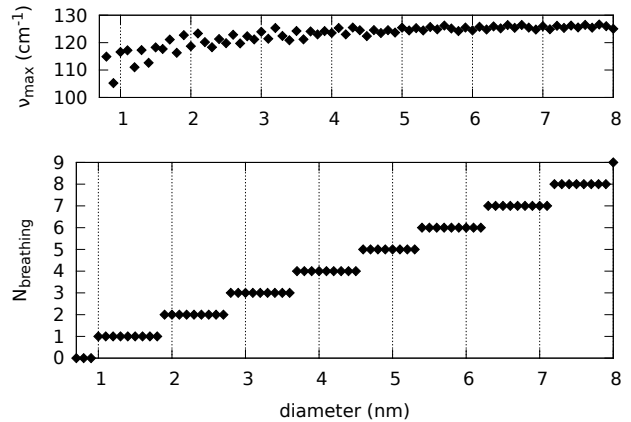}
\caption{\label{lamb}Top: variation of the highest frequency of the acoustic vibrations of a gold NC as a function of its diameter.
Bottom: maximum number of breathing vibrations $S_0^n$ for a gold NC as a function of its diameter.}
\end{figure}

Fig.~\ref{lamb} shows $N_{breathing}$ and $\nu_{max}$ as a function of the NC diameter.
It appears that a gold NC can sustain an $S_0^n$ breathing mode only if the NC diameter is larger than 1.0, 1.9, 2.8, 3.7 and 4.6~nm for $n=1, 2, \ldots 5$ respectively.
Therefore, the number of observed Raman peaks in Fig.\ref{exp} is determined (at least partly) by the NC size.
For instance, the $S_0^4$ vibration mode is observed in the Raman spectrum of the $13$~nm NC (Fig.~\ref{exp}).
But such a mode is absent in the spectrum of $3.5$~nm NCs because it does not exist for this NC size (independently from the associated Raman scattering efficiency).  

Unlike $N_{breathing}$, $\nu_{max}$ weakly depends on the NC size and is between $100$ and $125$~cm$^{-1}$ for $d$ ranging from 1 to 8$~nm$.
For large diameters, this value is close to the Debye frequency for gold (118~cm$^{-1}$) due to the similar assumptions used in Lamb and Debye models.
Hence, the frequency range where the low-frequency Raman peaks are expected weakly depends on the NC size. It certainly cannot reach very high frequencies for small sizes.\cite{KurbatovaPSS10}

\subsection{Atomistic approach}

The previous discussion provides a basic approach to the understanding of the number of vibration modes a NC can sustain.
However, since this approach is based on the assumption of a continuous and isotropic elastic medium, it fails to correctly describe the vibration dynamics of surface atoms~\cite{Combe2009} and the non-linearity of the phonon dispersions.
As a matter of fact, in bulk gold the phonon dispersion significantly deviates from the linear variation
around 100 and 60~cm$^{-1}$ for longitudinal and transverse phonons, respectively.\cite{LynnPRB73}

Atomistic calculations have been carried out using the same method and parameters
as those reported in Ref.~\onlinecite{CombePRB09}.
The size imposed limitation of allowed vibration modes was already pointed out for the $S_0^2$ mode.\cite{CombePRB09}
In the following, we shall focus on the overtones $S_0^n$ of the breathing mode for the NC size $d \simeq  3.5$~nm investigated experimentally in this work.
Calculations were performed for gold NCs with 755, 959 and 1270 atoms corresponding to $d \simeq 3.16$, $3.34$ and $3.7$~nm, respectively.
We obtain $\nu_{max}\simeq150$~cm$^{-1}$ a value close to the one estimated using Lamb's model (Fig.~\ref{lamb}).


\begin{figure}
\includegraphics[width=\columnwidth]{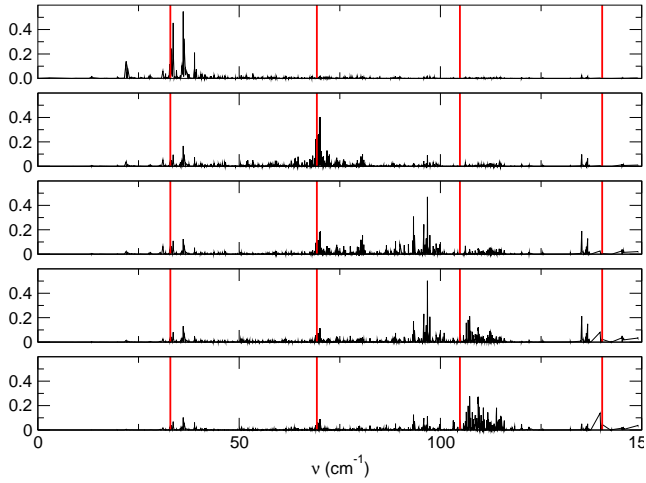}
\caption{\label{gold755}(Color online) Projection of the overtones of the breathing vibration modes $S_0^n$ with $n=1$ to 5 from top to bottom  on the atomistic eigenmodes of a gold NC with 755 atoms ($d \simeq 3.16$~nm) as a function of the atomistic eigenfrequencies. The frequencies of the $S_0^n$ vibrations are indicated by vertical lines (red online) where $n=1, \ldots 4$ from left to right.}
\end{figure}

\begin{figure}
\includegraphics[width=\columnwidth]{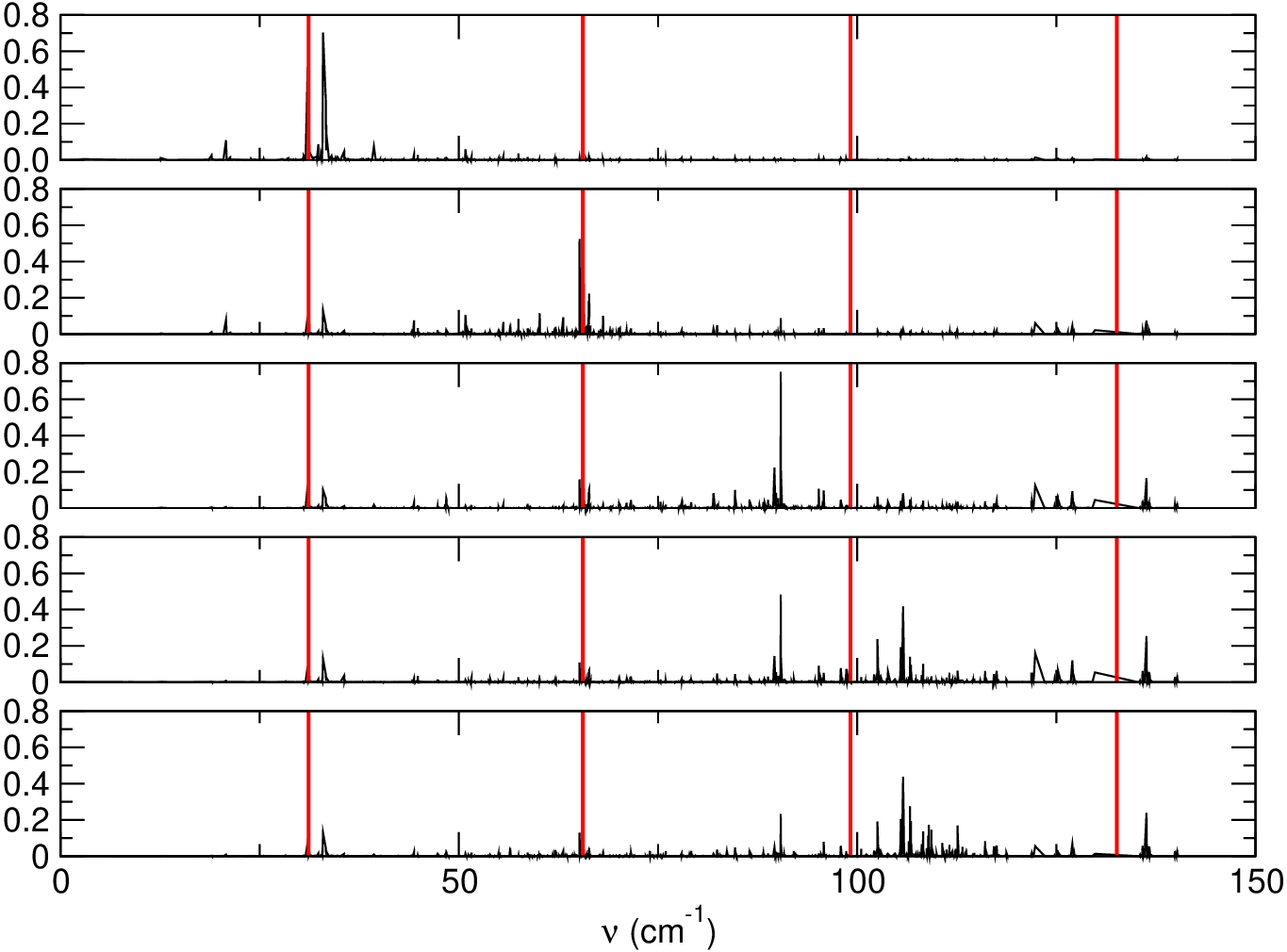}
\caption{\label{959gold}(Color online) Same as Fig.~\ref{gold755} for a gold NC with 959 atoms ($d \simeq 3.34$~nm).}
\end{figure}

\begin{figure}
\includegraphics[width=\columnwidth]{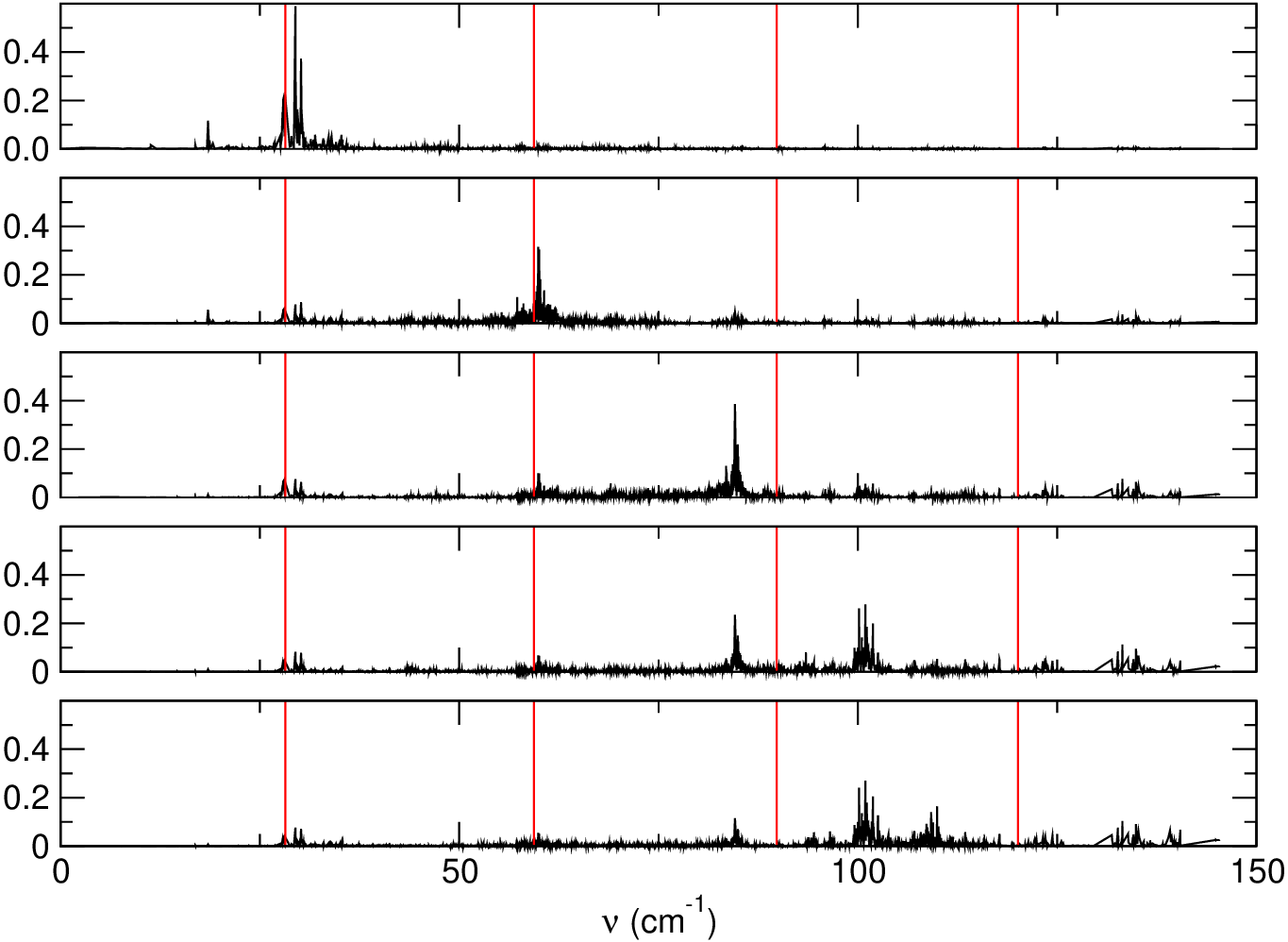}
\caption{\label{gold1270}(Color online) Same as Fig.~\ref{gold755} for a gold NC with 1270 atoms ($d \simeq 3.7$~nm).}
\end{figure}

Following Ref.~\onlinecite{CombePRB09}, the atomistic eigenmodes have been projected onto Lamb modes $S_0^n$ in order to evidence some of the atomistic vibrations which may contribute to the Raman spectra.
Figs.~\ref{gold755}, \ref{959gold} and \ref{gold1270} show such projections for NC with 755, 959 and 1270 atoms.
Due to the orthogonality and normalization of the atomistic modes, the sum of the squared projections onto a given Lamb mode is equal to one.

The $S_0^1$ and $S_0^2$ projection peaks are very close to the frequencies of the $S_0^1$ and $S_0^2$ Lamb modes.
This points out the good agreement between the vibration frequencies calculated using the atomistic model and Lamb's model.
The frequency of the $S_0^3$ projection peak is significantly lower than that of the $S_0^3$ Lamb mode, mainly because of the non-linearity of the phonon dispersion not accounted for in the continuous medium approximation.
In addition, significant projections are observed in a broad frequency range, especially for the smallest NC size (755 atoms, Fig.~\ref{gold755}).
Significant projections occur also at the frequencies of the $S_0^1$ and $S_0^2$ Lamb modes.
This is even more pronounced for the projections onto the $S_0^4$ and $S_0^5$ Lamb modes (Figs.~\ref{gold755}, \ref{959gold} and \ref{gold1270}), which indicates that the atomistic eigenmodes strongly differ from the $S_0^4$ and $S_0^5$ Lamb modes.

\begin{figure}
\includegraphics[width=\columnwidth]{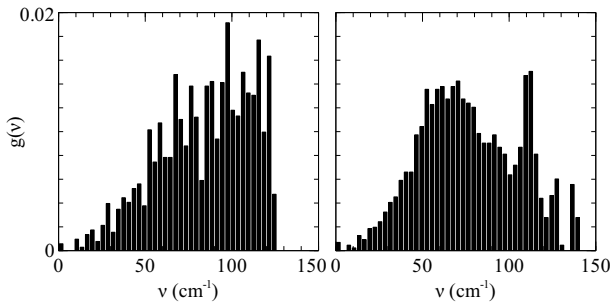}
\caption{\label{vdos}(Color online) VDOS for a 3.34~nm (959 atoms) Au NC calculated from Lamb's model (left) and the atomistic model (right). The bin size is 3~cm$^{-1}$.}
\end{figure}

Fig.~\ref{vdos} shows the vibrational density of states (VDOS) calculated using the atomistic model and Lamb model. For the latter, the size-imposed restriction of the number of allowed vibration modes has been taken into account as well as the $2 \ell + 1$ degeneracy of the spheroidal and torsional modes. 
The shape of the atomistic VDOS is similar to the one already reported for bulk gold,\cite{LynnPRB73} whereas the Lamb VDOS is quite different. First, it is not smooth especially at high frequency due to the large degeneracy of the involved modes.
Second there is no high frequency tail but rather an abrupt cut-off (due to the size-limited number of modes).
It clearly appears that Lamb model is not reliable in the high frequency range of the acoustic spectrum.

\begin{figure}
\includegraphics[width=0.75\columnwidth]{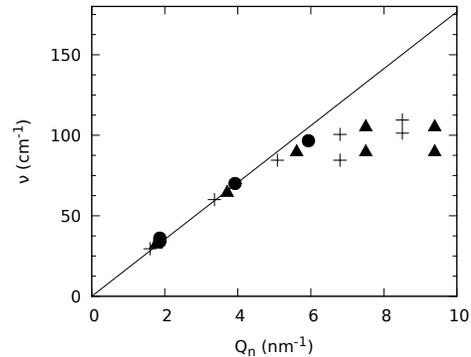}
\caption{\label{disp} (Color online) Frequencies of the $n$-breathing-like atomistic(red) and Lamb $S_0^n$ (black) modes as a function of $Q_n$.
The number of atoms in the Au NC is 755, 959 and 1270 for the circles, triangles and crosses respectively.
}
\end{figure}

In order to investigate more precisely the validity of Lamb model, we plot in Fig.~\ref{disp} the frequencies of the $n$-breathing-like atomistic modes as a function of a pseudo-wave vector $Q_n$.
We define the $n$-breathing-like atomistic modes as the modes with projections on Lamb modes $S_0^n$ greater than 75\% of the maximum of these projections (see Figs.~\ref{gold755}, \ref{959gold} and \ref{gold1270}).
To these modes, we associate a pseudo-wave vector $Q_n = \frac{2 \pi \nu_{S_0^ n}}{v_l}$ that accounts for the radial dependence of their displacement field; $ \nu_{S_0^ n}$ being the frequency of the $S_0^n$ mode and $v_l$ the longitudinal sound speed in gold.
In Lamb's theory, the displacement field of the $S_0^n$ Lamb mode is proportional to $j_1(Q_n r)$; $j_1$ being the spherical Bessel function of order 1.\cite{lamb1882}

Fig.~\ref{disp} shows the frequencies of the $n$-breathing-like atomistic modes as a function of $Q_n$. They are compared to the linear dependence of the Lamb frequencies $\nu_{S_0^ n}$. 
Despite a relative dispersion of the atomistic frequencies (especially for $Q_n \gtrsim 5$~nm$^{-1}$) due to the uneven projections, the frequencies of the atomistic modes obtained for the different NC sizes approximately overlap.
As can be seen, for $Q_n \lesssim 5$~nm$^{-1}$, the vibration frequencies obtained using the atomistic and Lamb models are in a very good agreement.
However, for $Q_n \gtrsim 5$~nm$^{-1}$, Lamb model clearly overestimates the vibration frequencies in comparison with the atomistic model.
Since $Q_n$ is a rough estimation of a pseudo-wave vector for the spherical symmetry, the plots in Fig.~\ref{disp} reflect the phonon dispersion curves for both models and therefore point out the difference between the atomistic calculations and Lamb model.

\subsection{Comparison with experiments}

\subsubsection{Raman scattering}

The spectra shown in Fig.~\ref{exp} clearly demonstrate the main result of our calculations, namely that the number of observable Raman peaks is determined by the NC size.
However, one must first check that the inhomogeneous broadening due to the size distribution of the NCs does not prevent the observation of some Raman features.

Assuming that the linewidth of the Raman lines is due only to inhomegeneous broadening (\textit{i.e.}, neglecting the homogeneous contribution) and that the scattered intensity for each peak does not depend on size, the linewidth/frequency ratio is the same for all the Raman peaks.
This ratio is $r \simeq 0.35$ for the $S_2^1$ $E_g$ peak of the 3.5~nm NCs and $r \simeq 0.25$ for the $S_0^1$ peak of the 13~nm NCs.
Two consecutive overtones will significantly overlap when their frequency separation becomes equal to their full width at half maximum.
Because $\omega_{S_0^n} \simeq n \omega_{S_0^1}$, this occurs for $n \gtrsim 1/r$ \textit{i.e.}, $n \gtrsim 3$ and 4 for the 3.5 and 13 nm NCs, respectively.
If the homogeneous broadening is not negligible (which is the case in this work since the width of the size distribution is about 15\% of the average size for both samples) or if the scattered intensity for a given peak depends strongly on the size then the linewidth increases more slowly with $n$ and more overtones can be observed.
Therefore the size distribution does not explain the non-observation of the $S_0^2$ peak in the spectrum of the smallest NCs. However it may be responsible for the non-observation of the $S_0^6$ peak of the 13 nm NCs.

The Raman scattering efficiency decreases with increasing overtones\cite{BachelierPRB04} which makes the observation of such high order modes ($n>1$) very challenging.\cite{TripathyNL11}
The Raman spectrum of the 3.5~nm NCs (Fig~\ref{exp}) does not evidence any peak corresponding to the first harmonic of the breathing mode ($S_0^2$) despite the resonance and interference enhancement.
However the atomistic (Figs.~\ref{gold755}, \ref{959gold}, \ref{gold1270}) and Lamb (Fig~\ref{lamb}) models point to the existence of such a mode.
Modeling of the Raman scattering efficiencies based on an atomistic description of the NC vibration eigenmodes and their interactions with the electronic states is required in order to more precisely address this issue.\cite{Combe2007}

\subsubsection{Transient absorption}
 
The breathing modes of gold NCs can also be observed using time-resolved transient absorption experiments. Up to now, fewer overtones of the breathing vibrations have been reported using this technique compared to Raman scattering.\cite{NeletASS04,TripathyNL11}
Recently, Juve \textit{et al.}\cite{Juve10} conducted time-resolved transient absorption experiments on small ($1.3 < d < 3$~nm ) platinum NCs and investigated the size-dependence of the fundamental $S_0^1$ breathing modes: no discrepancy between the measured vibration frequencies and the frequencies calculated using the Lamb model has been found for NCs as small as 1.3~nm. 

According to our atomistic calculations (Fig.~\ref{disp}), a strong deviation of Lamb frequencies from the atomistic frequencies is expected for $Q_n  \gtrsim 5$~nm$^{-1}$.
Assuming that the elastic properties of platinium and gold are similar, $Q_n  \gtrsim 5$~nm$^{-1}$ corresponds, for the fundamental breathing mode $n=1$, to a NC diameter $d \lesssim 1.2$~ nm. Below this critical size, Lamb model fails to describe the vibration dynamics of the NCs.
The minimum size (1.3~nm) investigated by Juve \textit{et al.}\cite{Juve10} is close but still above the critical size, thus explaining the good agreement between the measured frequencies and those estimated using Lamb model.

Time-resolved transient absorption experiments~\cite{VarnavskiACSNANO11} performed on very small Au NCs ($1.1 < d < 2.2$~nm) have evidenced an acoustic vibration mode, apparently the  $S_0^1$ breathing mode, whose frequency does not change with the NC size.
This behavior is consistent with our atomistic model and in particular with Fig.~\ref{disp}.
The largest NC size ($d=2.2$~nm) for which this frequency independent mode is observed\cite{VarnavskiACSNANO11} is in quite good agreement with our predictions.
However, the frequency (around 70~cm$^{-1}$) measured for the smallest NCs is smaller than the one expected from the atomistic model.
Note that this value is close to the first peak in the atomistic VDOS as pointed in Ref.~\onlinecite{VarnavskiACSNANO11}, suggesting that the observed mode may not be the $S_0^1$ breathing mode.
This again points to the need for an improved model based on an atomistic description of the NC and an accurate model of the electron-vibration coupling. 

\subsection{Other nanoparticles}

While this work was focused on gold NCs, similar results are expected for other metallic and semiconductor NCs.
However, a special feature of gold is the large longitudinal/transverse sound velocity ratio.
The frequency difference between two consecutive breathing modes is proportional to $v_L$ while this difference for other vibrations symmetry ($S_{\ell \neq 0}^ n$ and $T_\ell^n$) is mainly proportional to $v_T$.
Hence, there are more modes having a lower frequency than an $S_0^n$ vibration mode in gold than in most of the other materials.
For example, the number of such modes is about two times larger for gold than for silver and about three time larger than for copper.
Hence the impact of the size-induced reduction of the number of $S_0^n$ vibration eigenmodes is more easily evidenced in the case of gold NCs.

\section{Conclusion}

The number of peaks in the inelastic light scattering spectra of small gold NCs is shown to depend on the NC size.
This result can be qualitatively explained by restricting the number of vibration eigenmodes according to simple rules based on the number of atoms in the NC.
This effect comes in addition to the well-known $1/d$ dependence of the acoustic vibration frequencies.
Atomistic calculations provide a more exact but also more complex picture than Lamb's model.
Complexity arises from the non-linearity of the phonon dispersion curves which is very pronounced for vibration modes with wavelengths comparable to the interatomic distance.
By comparing the sequence of frequencies obtained from the atomistic and Lamb's models, we were able to define a frequency above which the continuous medium approximation is no longer valid.
This frequency is neither size- nor mode-dependent.
It is mainly determined by the non-linearity of the phonon dispersions.

\begin{acknowledgments}
L. S. acknowledges D. B. Murray for his critical reading of the manuscript.
\end{acknowledgments}

\bibliography{nldc}
\end{document}